\documentclass[10pt,conference]{IEEEtran}
\usepackage[final]{pdfpages}

\usepackage[a4paper, top=0.85in, bottom=1.05in, left=1.2cm, right=1.2cm]{geometry}

\ifCLASSINFOpdf
 \else
\fi

\usepackage{wrapfig}
\usepackage{comment}
\usepackage{balance}  
\usepackage{cite}
\usepackage{graphicx}
\usepackage{float}
\usepackage{algorithm}
\usepackage{algpseudocode}
\usepackage{amsmath} 
\usepackage{cuted} 
\usepackage[utf8]{inputenc}
\usepackage[final]{pdfpages} 
\usepackage{pdfpages}
\usepackage{lipsum}
\usepackage{textcase}
\usepackage{url}
\usepackage{amsmath,esint}
\usepackage{epstopdf}
\usepackage{array} 
\usepackage{multirow}
\usepackage{booktabs}
\usepackage{esvect}
\usepackage{soul}
\usepackage[normalem]{ulem} 
\usepackage{balance}
\usepackage{multicol}

\usepackage{subcaption}         

\usepackage[justification=centering]{caption}

\newcolumntype{P}[1]{>{\centering\arraybackslash}p{#1}}
\newcolumntype{M}[1]{>{\centering\arraybackslash}m{#1}}

\begin{document}


\title{Predicting Lifespan of Ground-to-Air Multipath Components in mmWave UAV Channels}

\author{\IEEEauthorblockN{Wahab Khawaja\IEEEauthorrefmark{1},~Rune H. Jacobsen\IEEEauthorrefmark{1}, 
Sajid Hussain\IEEEauthorrefmark{2},~ and Ismail Guvenc\IEEEauthorrefmark{3}
}

\IEEEauthorblockA{\IEEEauthorrefmark{1}Dept. Electrical and Computer Engineering, Aarhus University, 8000 Aarhus, Denmark}

\IEEEauthorblockA{\IEEEauthorrefmark{2}Dept. Computer Systems Engineering, Mirpur Univ. Science and Technology, Mirpur AJK, 10250, Pakistan}

\IEEEauthorblockA{\IEEEauthorrefmark{4} Dept. Electrical and Computer Engineering, North Carolina State University, Raleigh, NC 27606, USA}

Email: wahabgulzar@ece.au.dk, rhj@ece.au.dk, engrsajid.cse@must.edu.pk, iguvenc@ncsu.edu
}

\maketitle
\thispagestyle{empty}
\begin{abstract}
In mobile ground-to-air (GA) propagation channels, the birth and death of multipath components (MPCs) are frequently observed, and the wide-sense stationary uncorrelated scattering (WSSUS) assumption does not always hold. Several methods exist for tracking the birth and death of MPCs, however, to the best of knowledge of authors, there is no existing literature that addresses the prediction of the lifespan of the MPCs in non-WSSUS GA propagation channels. In this work, we consider the GA channel as non-WSSUS and individual MPCs across receiver positions are represented as time series based on the Euclidean distance between channel parameters of the MPCs. These time series representations, referred to as path bins, are analyzed using a semi-Markov chain model. The channel parameter variations and dependencies between path bins are used to predict the lifespan of path bins using weighted sum method, machine learning classifiers, and deep neural networks. For comparison, the birth and death of path bins are also modeled using a Poisson distribution and a Markov chain. Simulation results demonstrate that deep neural networks offer highly accurate predictions for the lifespan~(including death) of MPC path bins in the considered GA propagation scenario. 
\end{abstract}


\IEEEpeerreviewmaketitle

\section{Introduction}
Ground-to-air~(GA) communications using unmanned aerial vehicles~(UAVs) have secured significant attention in the recent decade~\cite{uav_AG_ref1, uav_AG_ref2}. GA communications using UAVs are envisioned for $5$G and beyond~\cite{uav_5G_beyond1,uav_5G_beyond2}. The integration of 5G terrestrial networks with non-terrestrial networks~(NTN) including UAVs is highlighted in the latest release of the 3GPP~\cite{3GPP_ntn}. The 5G and beyond communication networks are expected to dominantly use millimeter wave~(mmWave) frequencies. In the mmWave mobile GA propagation channel, there are sparse multipath components~(MPCs) that experience frequent births and deaths~\cite{birth_death1,birth_death2,wahab_AG_previous}.


Accurate prediction of lifespan of MPCs in GA channels is essential for maintaining reliable links and optimizing resources in NTN. This enables better handovers and minimizes link failures under dynamic conditions. There are different methods available in the literature for tracking the lifespan of MPCs including their birth and death~\cite{birth_death_tracking1,birth_death_tracking2,birth_death_tracking4}. In \cite{birth_death_tracking1}, nonlinearities due to the birth and death of MPCs in a vehicular propagation channel were detected and tracked using a modified Kalman filter. In \cite{birth_death_tracking2}, the analysis of the birth and death of MPCs in a time-variant channel was carried out using spatial point process. Assumption of wide sense stationary channel transfer function using point process perspective helped in the analysis. In \cite{birth_death_tracking4}, a statistical model for time-varying UAV air-to-ground~(AG) channel was presented. The stochastic model provided characterization of the parameters of the birth and life of MPCs. Furthermore, it was shown that regular patterns can be used to describe the evolution of MPCs. Birth and death of MPCs was also analyzed using AG propagation channel measurements in~\cite{BD_Matolak}. However, to the best of knowledge of authors, no existing study has focused on predicting the lifespan of MPCs in mobile GA propagation channels, motivating this research to bridge that gap. This work highlights the need to address the non-WSSUS nature of GA channels in vehicular communications~\cite{non-wssus}, where UAV motion induces non-WSS, and static scatterers create correlated MPCs. 

\begin{figure*}[ht]
    \centering
    \begin{minipage}{0.32\textwidth}
        \centering
        \includegraphics[width=\linewidth]{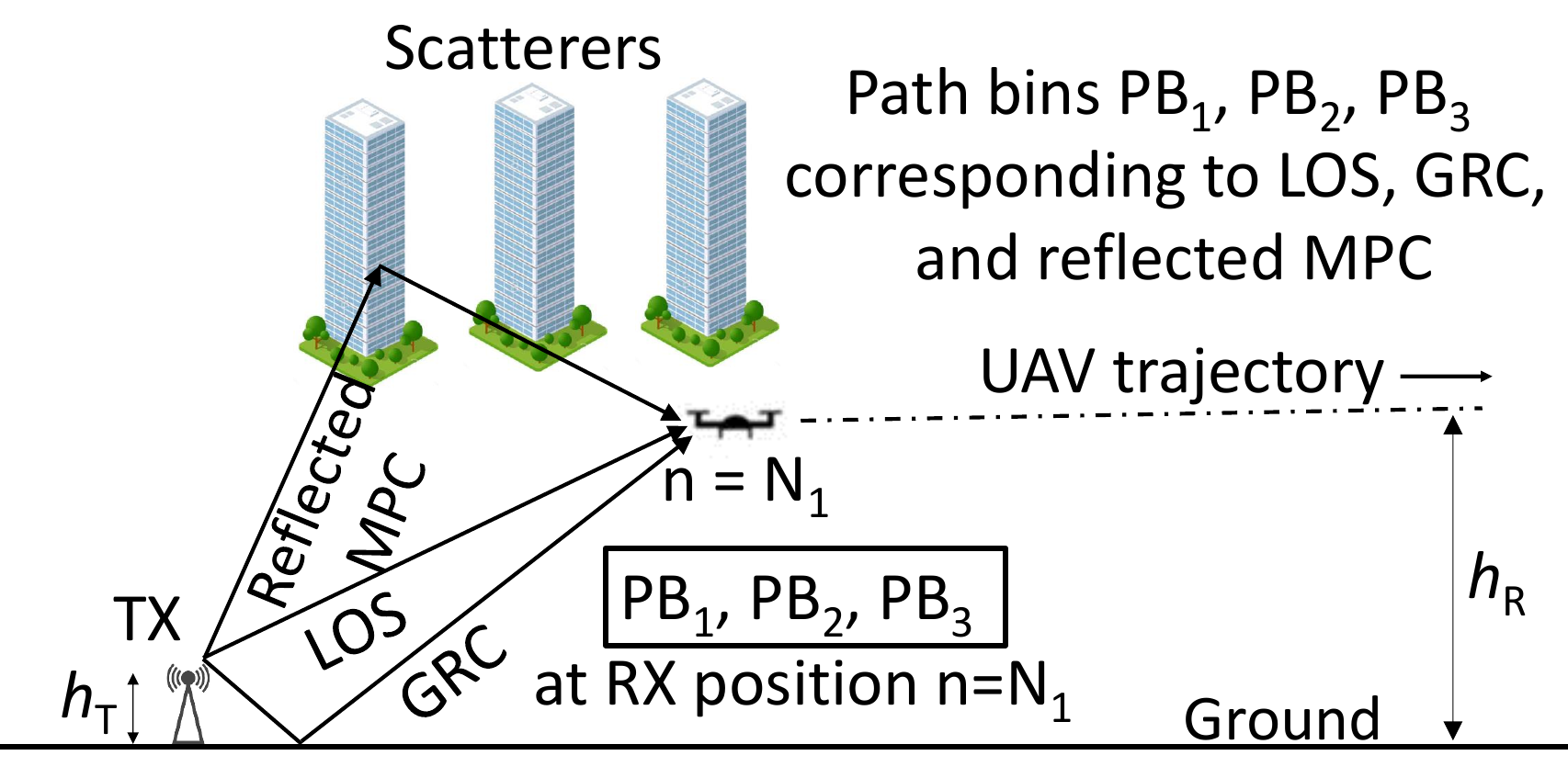}
        \subcaption{}
        \label{fig:fig1}
    \end{minipage}%
    \hfill
    \begin{minipage}{0.32\textwidth}
        \centering
        \includegraphics[width=\linewidth]{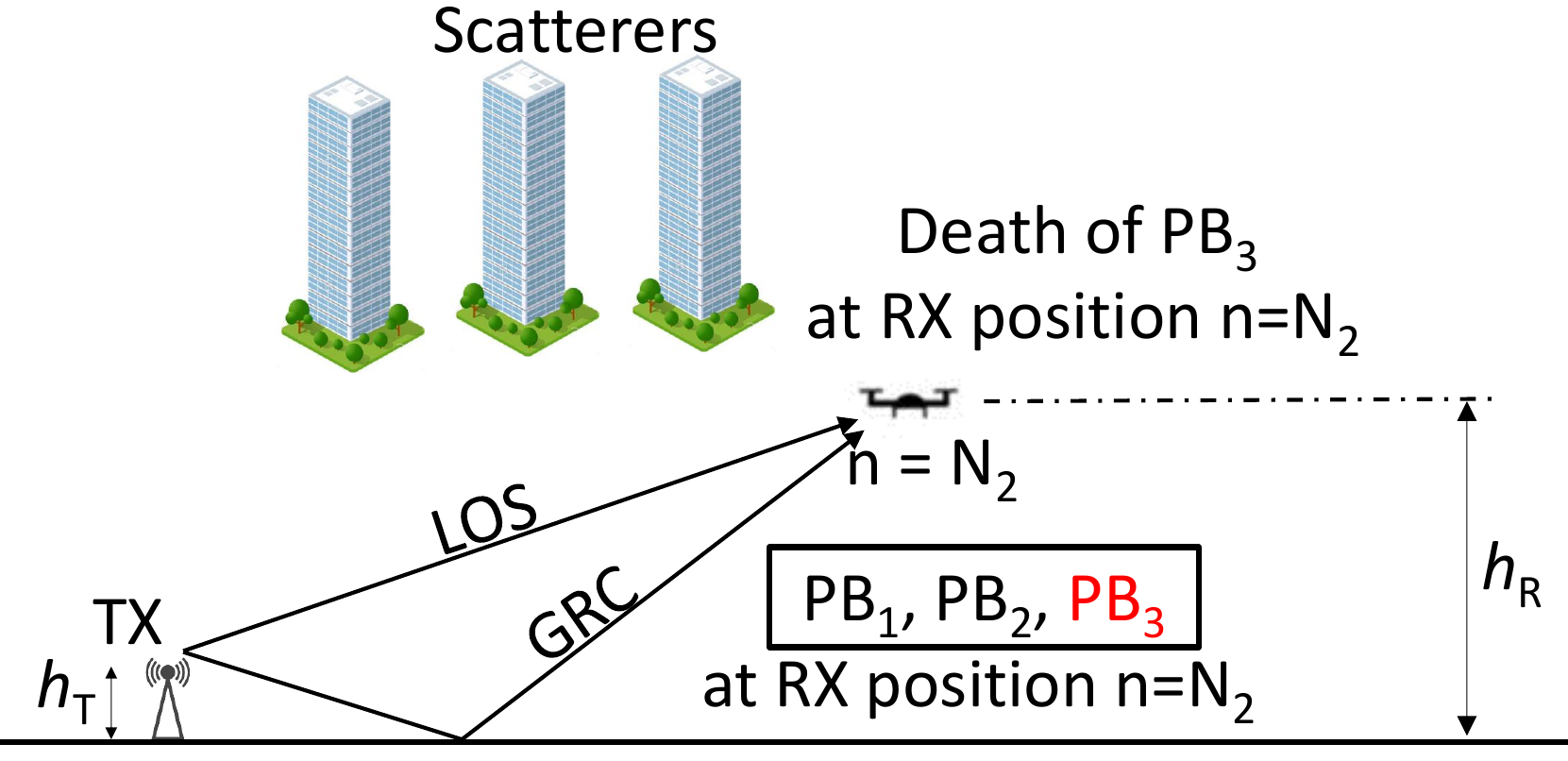}
        \subcaption{}
        \label{fig:fig2}
    \end{minipage}%
    \hfill
    \begin{minipage}{0.32\textwidth}
        \centering
        \includegraphics[width=\linewidth]{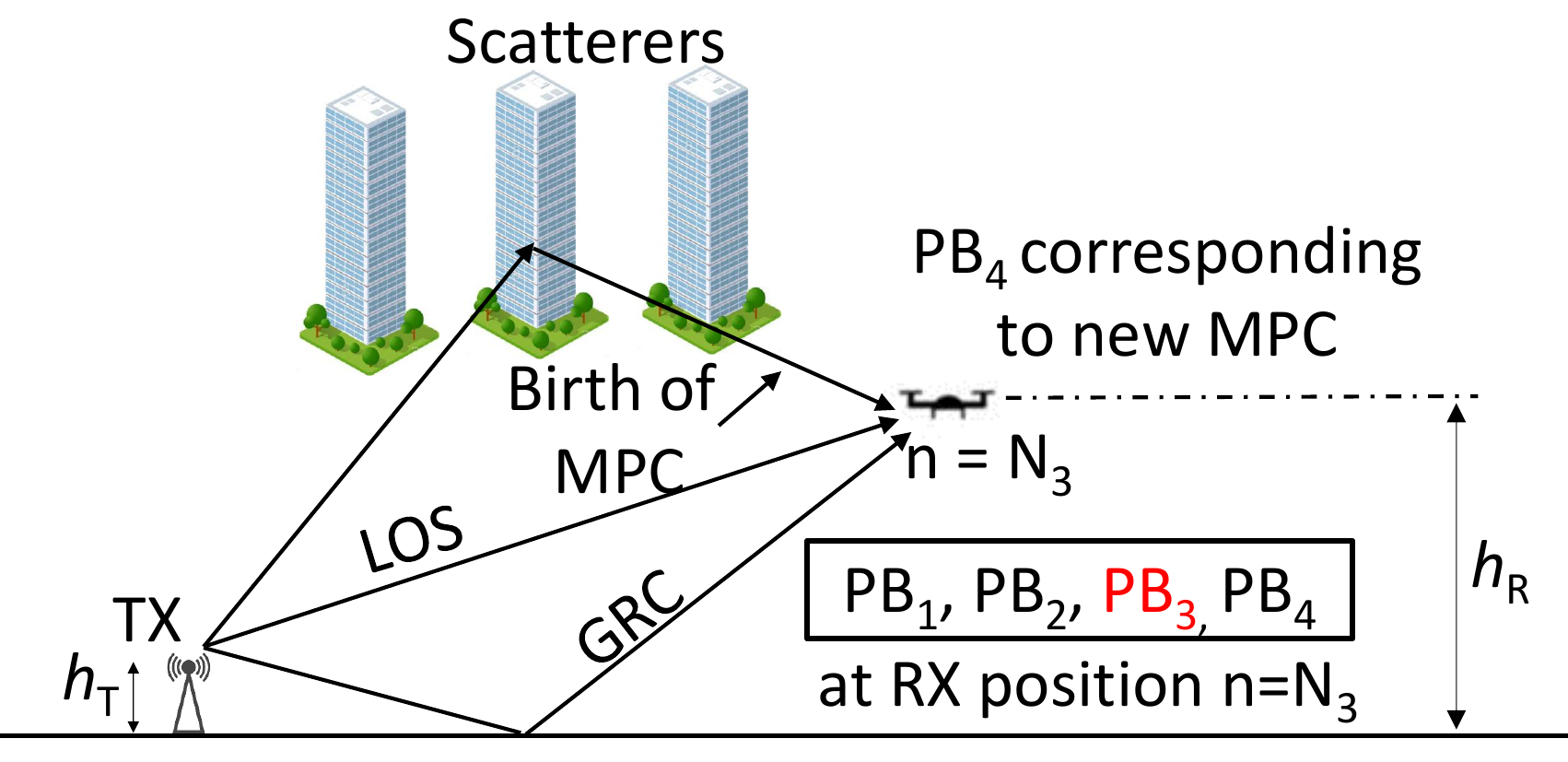}
        \subcaption{}
        \label{fig:fig3}
    \end{minipage}
    \caption{The GA propagation scenario showing life span, and birth and death instances of MPCs along the UAV trajectory.}
    \label{Fig:LOS_Refl_diff}
\end{figure*}

The contributions of this work are as follows: We analyze a mobile GA directional propagation channel with scatterers positioned on one side shown in Fig.~\ref{Fig:LOS_Refl_diff}. First, the tracking of the birth and death of MPCs is obtained by organizing the MPCs as respective time series over receiver~(RX) positions. The time series representation of a MPC called the path bin is obtained using the Euclidean distance among the channel parameters of MPCs at different RX positions\cite{wahab_AG_previous} highlighted in Fig.~\ref{Fig:LOS_Refl_diff}. The lifespan of path bins defined as the number of RX positions covered until their death is predicted using predictive modeling based on the analysis of the characteristics of the path bins and the correlation among the path bins. Weighted sum~(WS) models, machine learning~(ML), and deep neural networks~(DNN) are used to predict the lifespan of path bins. The birth and death of path bins are also modeled as independent processes using Poisson distribution and Markov chain. The prediction performance of different methods is then compared. Geometric ray-based method is used for simulations. 

The outline of the paper is as follows, Section~\ref{Section:TS_represent} provides the time series representation of MPCs as path bins over the RX positions, Section~\ref{Section:birth_death_model} provides different methods for the prediction of lifespan of MPC path bins, the simulation setup and results are discussed in Section~\ref{Section:simulations}, and Section~\ref{Section:conclusions} concludes the paper. 

\section{Time Series Representation of MPCs}  \label{Section:TS_represent}
We consider our GA propagation channel shown in Fig.~\ref{Fig:LOS_Refl_diff} to be a time-varying channel with correlated scattering. We model our channel $\pmb{H(t,n)}$ as follows:
\begin{align}
\pmb{H(t,n)} &=  \sum_{m=1}^{M_n}\alpha_{m,n}(t)\delta \big(t - \tau_{m,n}\big) \delta \big(\pmb{\theta^{(\rm T)}} - \theta_{m,n}^{(\rm T)}\big)  \nonumber \\ 
&\times \delta \big(\pmb{\phi^{(\rm T)}} - \phi_{m,n}^{(\rm T)}\big)\delta \big(\pmb{\theta^{(\rm R)}} - \theta_{m,n}^{(\rm R)}\big) \delta \big(\pmb{\phi^{(\rm R)}} - \phi_{m,n}^{(\rm R)}\big)\nonumber \\
& \times \exp{\bigg(\frac{-j2\pi \Delta d_{m,n}}{\lambda}\bigg)}~,
\end{align} 
where $H(t,n)$ is the channel impulse response at time instance $t$ at $n^{\rm th}$ RX position, $\alpha_{m,n}$ and $\tau_{m,n}$ are the complex amplitude and delay of the $m^{\rm th}$ MPC at $n^{\rm th}$ RX position, respectively, $\pmb{\theta^{(\rm T)}}$ and $\pmb{\phi^{(\rm T)}}$ are the elevation and azimuth angle vectors at the transmitter~(TX), respectively, $\pmb{\theta^{(\rm R)}}$, $\pmb{\phi^{(\rm R)}}$ are the elevation and azimuth angle vectors, respectively, at the RX, $\theta_{m,n}^{(\rm T)}$, $\phi_{m,n}^{(\rm T)}$ are the elevation and the azimuth angles of the $m^{\rm th}$ MPC corresponding to $n^{\rm th}$ RX position at the TX, respectively, $\theta_{m,n}^{(\rm R)}$, $\phi_{m,n}^{(\rm R)}$, are the elevation and the azimuth angles of the $m^{\rm th}$ MPC corresponding to $n^{\rm th}$ RX position at the RX, respectively, $n=1,2,\cdots,N$ and $M_n$ are the total number of RX positions and MPCs at $n^{\rm th}$ RX, respectively, and $\Delta d_{m,n} = d_{m,n} - d_0$, where $d_{m,n}$ is the distance of the $m^{\rm th}$ MPCs at $n^{\rm th}$ RX position, $d_0$ is the reference distance. The phase of the MPCs $\big(\frac{2\pi \Delta d_{m,n}}{\lambda}\big)$ at different RX positions are assumed to be uniformly distributed between [0$-$2$\pi$). 


The MPCs observed at the RX exhibit strong correlations because of dependencies in channel parameters across RX positions along the straight-line trajectory. The predictable variation in MPC parameters, caused by consistent changes in distance to scatterers, supports effective prediction of MPC lifespan through Euclidean distance and correlation-based methods. The correlation among the channel parameters of MPCs at $n^{\rm th}$ and $k^{\rm th}$ RX positions can be represented by the matrix $\pmb{R}$ defined as
\small
\begin{equation}
\pmb{R} = 
\begin{bmatrix}
{\textbf{c}}_{1,1}&{\textbf{c}}_{1,2}&{\textbf{c}}_{1,3}&\cdots & {\textbf{c}}_{1,M_k}\\
{\textbf{c}}_{2,1}&{\textbf{c}}_{2,2}&{\textbf{c}}_{2,3}&\cdots & {\textbf{c}}_{2,M_k}\\
\vdots & \vdots &\vdots &\cdots& \vdots\\
{\textbf{c}}_{M_n,1}&{\textbf{c}}_{M_n,2}&{\textbf{c}}_{M_n,3}&\cdots & {\textbf{c}}_{M_n,M_k}
\end{bmatrix}~,  \nonumber
\end{equation}
\normalsize
where ${\textbf{c}}_{m_n,m_k}~{\rm for}~ n\neq k$ is the correlation vector representing correlation among the channel parameters of the MPCs, $m_n$ and $m_k$ at the $n^{\rm th}$ and $k^{\rm th}$ RX positions, respectively. This vector is given by
\begin{align}
    &{\textbf{c}}_{m_n,m_k} = \nonumber \\ &\big[c_{m_n,m_k}^{(\alpha)},~ c_{m_n,m_k}^{(\tau)},~ c_{m_n,m_k}^{(\theta^{(\rm T)})},~ c_{m_n,m_k}^{(\phi^{(\rm T)})},~ c_{m_n,m_k}^{(\theta^{(\rm R)})},~ c_{m_n,m_k}^{(\phi^{(\rm R)})}\big],
\end{align}
where 
$c_{m_n,m_k}^{(\alpha)},~ c_{m_n,m_k}^{(\tau)},~ c_{m_n,m_k}^{(\theta^{(\rm T)})},~ c_{m_n,m_k}^{(\phi^{(\rm T)})},~ c_{m_n,m_k}^{(\theta^{(\rm R)})},~ c_{m_n,m_k}^{(\phi^{(\rm R)})}$
are the correlations between the $m_n^{\rm th}$ and $m_k^{\rm th}$ MPC's complex amplitude, delay, TX elevation and azimuth angles, and RX elevation and azimuth angles, respectively. 

\begin{figure*}[ht]
    \centering
    \begin{subfigure}{0.49\textwidth}
        \centering
        \includegraphics[width=\textwidth]{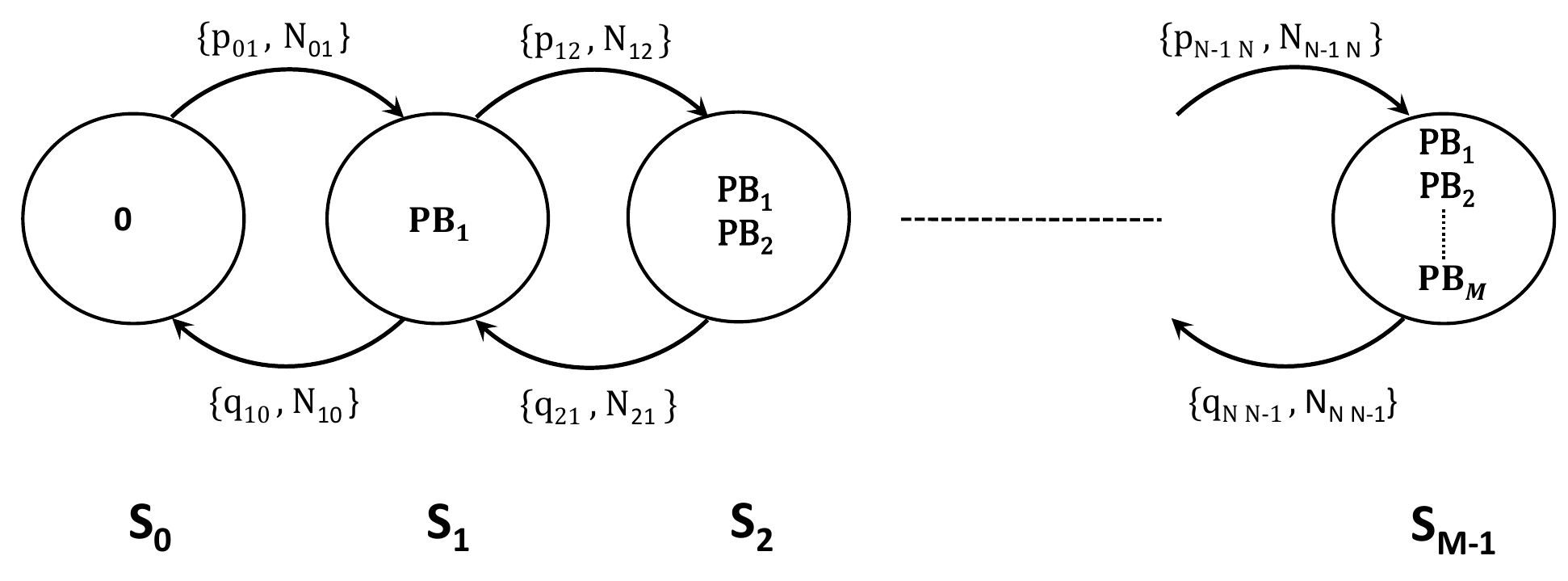}
        \caption{}
        \label{fig:figure2}
    \end{subfigure}
    \begin{subfigure}{0.49\textwidth}
        \centering
        \includegraphics[width=\textwidth]{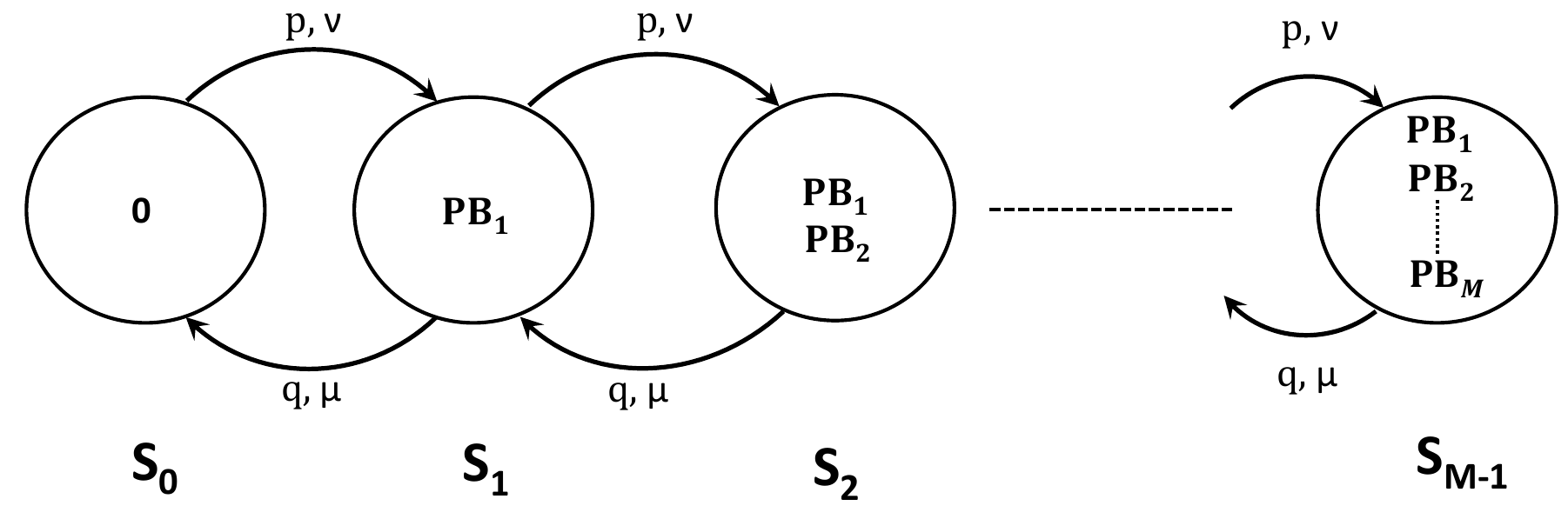} 
        \caption{}
        \label{fig:figure1}
    \end{subfigure}
    \hfill
        \captionsetup{justification=raggedright, singlelinecheck=false} 
    \caption{Birth and death modeling of path bins using (a) Semi-Markov chain for the WS, ML, and DNN methods; and (b) Markov chain for the Poisson distribution. }
    \label{Fig:Markov_chain}
\end{figure*}


 Correlations among MPC channel parameters at different RX locations enable path bin creation using Euclidean distance, grouping MPCs based on proximity in parameter space. Each MPC at a given RX position is assigned to a path bin by evaluating its channel parameter distance to all prior MPCs. Let ${\textbf{MC}}_{m,n}$ denote the $m^{\rm th}$ MPC vector at the $n^{\rm th}$ RX position, represented by six channel parameters as:
 \begin{align}
     {\textbf{MC}}_{m,n} = \big(\alpha_{m,n},\tau_{m,n},\theta_{m,n}^{(\rm T)}, \phi_{m,n}^{(\rm T)}, \theta_{m,n}^{(\rm R)}, \phi_{m,n}^{(\rm R)}\big). \label{Eq:ch_par}
 \end{align}
 The Euclidean distance between the $m^{\rm th}$ MPC at $n^{\rm th}$ RX position and $j^{\rm th}$ MPC at $k = n-1,n-2,\cdots,2,1$, RX positions is given as 
\begin{align}
d^{(\rm u)}(m,n,j,k) = {\rm Euclidean}\big({\textbf{MC}}_{m,n}(u),{\textbf{MC}}_{j,k}(u)\big), 
\end{align}
where $u={1,2,\cdots,6}$ are the channel parameters given in (\ref{Eq:ch_par}). The total distance between the $m^{\rm th}$ MPC at $n^{\rm th}$ RX position and $j^{\rm th}$ MPC at $k = n-1,n-2,\cdots,2,1$, RX positions is 
\begin{align}
    d(m,n,j,k) = \frac{1}{\gamma}\sum_{u=1}^{6}d^{(\rm u)}(m,n,j,k),
\end{align}
where $\gamma$ is the normalizing factor. Our Euclidean distance-based method for grouping MPCs into path bins~\cite{wahab_AG_previous} is effective for straight-line and gently curved UAV trajectories, where gradual changes in channel parameters ensure predictable correlations. For complex paths, such as circular trajectories, cyclic dependencies can require adjustments to the distance metric to accurately capture MPC behavior. 

A path bin obtained using Euclidean distance represents a time series of an individual MPC over the RX positions. If ${\textbf{PB}}$ represents a path bin, an $m^{\rm th}$ path bin can be written as ${\textbf{PB}_m} =  {\textbf{MC}}_{m,N_m}$, where $N_m$ represent the lifespan of $m^{\rm th}$ MPC over RX positions. If there are $M$ path bins given as $\big[ {\textbf{PB}_1}~ {\textbf{PB}_2} ~\cdots~ {\textbf{PB}_M} \big]$ they can be represented as follows
 \small
\begin{equation}
\big[{\textbf{MC}}_{1,N_1}~ {\textbf{MC}}_{2,N_2}~ {\textbf{MC}}_{3,N_3}~ \cdots~  {\textbf{MC}}_{M,N_M}\big]. 
\end{equation}
\normalsize
The path bins spanning over multiple RX positions represent the lifespan of a given MPC is shown in Fig.~\ref{Fig:LOS_Refl_diff}. The figure also highlights birth and death instances of path bins.   

\section{Prediction of Lifespan of MPC Path Bins}    \label{Section:birth_death_model}

This section explores the modeling of birth and death of path bins using semi-Markov model and uses WS, ML classifiers and DNN to predict the lifespan of the path bins. The birth and death of path bins is also modeled using Poisson distribution and Markov chain.

\subsection{Using WS Method}
The correlation among the path bins across the RX positions in a non-WSSUS channel can be used to predict the future behavior of the path bins. A semi-Markov model can be used to illustrate the dependencies, as shown in Fig.~\ref{Fig:Markov_chain}(a). In this figure, the transition probabilities for the birth and death of path bins depend on the number of RX positions covered. The transition probability from state $i$ to $j$ is denoted by $p_{ij}$ for birth, and the probability from state $j$ to $i$ is denoted by $q_{ji}$ for death. The number of RX positions on which $p_{ij}$ and $q_{ji}$ depend are represented by $N_{ij}$ and $N_{ji}$, respectively. In addition to correlation among the path bins, the channel variations of the path bins across the RX positions help in the prediction of the lifespan of a path bin. By using the WS method, the lifespan of a path bin in terms of the number of RX positions can be modeled as
\begin{align}
    & w_0\Delta \alpha_m + w_1\Delta \tau_m + w_2\Delta \theta_m^{(\rm T)} + \nonumber \\
    & w_3\Delta \phi_m^{(\rm T)} + w_4\Delta \theta_m^{(\rm R)} + w_5\Delta \phi_m^{(\rm R)}+ \nonumber \\ 
    & w_6 C_m^{(\rm pos)} + w_7C_m^{(\rm neg)} + w_8(M_m^{(\rm oth)}-\overline{M}^{\rm (N)})   = N_{m}, \label{Eq:weight_ch_par}
\end{align}
where $\Delta \alpha_m$ is the variation in the amplitude, $\Delta \tau_m$ is the variation in the delay, $\Delta \theta_m^{(\rm T)}$ and $\Delta \phi_m^{(\rm T)}$ are the variations in the elevation and azimuth angles at the TX, respectively, the variations in the elevation and azimuth angles at the RX are represented, respectively, as $\Delta \theta_m^{(\rm R)}$ and $\Delta \phi_m^{(\rm R)}$, $C_m^{(\rm pos)}$ and $C_m^{(\rm neg)}$ are the number of positive and negative correlation coefficients of channel parameters of the $m^{\rm th}$ path bin with the other path bins, $M_m^{(\rm oth)}$ are the number of other path bins alive during the lifespan of the $m^{\rm th}$ path bin and $\overline{M}^{(\rm N)}$ is the mean number of path bins observed over all RX positions, and $w_0,w_1,\cdots,w_8$ are the weight coefficients. 

Over $M$ path bins, we can write (\ref{Eq:weight_ch_par}) in the matrix form as.
\begin{align}
    \mathbf{V} \boldsymbol{W} &= \mathbf{X} \label{eq:main}
\end{align}
where $\mathbf{V} =\bigg[\mathbf{m}_1^T~\mathbf{m}_2^T~\cdots  \cdots~\mathbf{m}_M^T\bigg]^T$, $\boldsymbol{W}=\bigg[w_o ~w_1 ~\cdots ~w_8\bigg]^T$, and $\mathbf{X}=\bigg[N_1~ N_2~ \cdots~ N_M\bigg]^T$, and
 each row vector \(\mathbf{m}_i\) for $i=1,2,\cdots,M$ is defined as:
\begin{align}
\mathbf{m}_i =
    &\bigg[\Delta \alpha_i, \Delta \tau_i, \Delta \theta_i^{(\rm T)}, \Delta \phi_i^{(\rm T)}, \Delta \theta_i^{(\rm R)}, \nonumber \\
    &\Delta \phi_i^{(\rm R)}, C_i^{(\rm pos)}, C_i^{(\rm neg)}, (M_i^{(\rm oth)} - \overline{M}^{(\rm N)})\bigg]~.
\end{align}
There are generally a large number of path bins compared to the weights, so the solution for matrices becomes over-determined. To find the weight matrix, we use normal equation solution~\cite{overdet_norm} as follows $\textbf{W} = \big(\textbf{V}^{T}\textbf{V}\big)^{-1}\textbf{V}^{T}\textbf{X}$. Optimum weights are obtained using Monte Carlo simulations. For an $i^{\rm th}$ path bin, ${\textbf{PB}}_{ i}$ that currently occupies $n$ RX positions, we use the weight matrix to predict its lifespan, $N^{(\rm prd)}$ as $N^{(\rm prd)} = \mathbf{m}_i\textbf{W}$. We will not predict the birth of a path bin using WS, ML and DNN methods.


\subsection{Using Machine Learning Classifiers}
To predict the lifespan of MPC path bins, we use ML classifiers trained on observed variations in channel parameters and correlations between path bins. Linear discriminant analysis~(LDA),
Naive Bayes~(NB), and random forest~(RF) classifiers are well-suited due to their ability to capture feature relationships: LDA handles linear separability well, NB is effective for probabilistic relationships, and RF manages complex patterns and correlations. The trained model detects parameter patterns associated with path bin death, allowing for lifespan predictions of new bins. Similar to the WS method, the transition probability for path bin death over a given number of RX positions, as shown in Fig.~\ref{Fig:Markov_chain}(a), can be derived through ML methods.

LDA, NB, and RF classifiers use (\ref{eq:main}) for predicting the lifespan of the path bins. The training data consists of matrix $\textbf{V}$ and matrix $\textbf{X}$ serves as the corresponding class labels. The model for classification and subsequent prediction of lifespan for an $i^{\rm th}$ test path bin is given as
\begin{equation}
\mathcal{C} = f^{(\rm ML)}\big(\textbf{V},\textbf{X}\big), ~ N^{(\rm prd)} = \rm{predict}(\mathcal{C},{\textbf{m}_i}),
\end{equation}
where $\mathcal{C}$ is the classifier model, $f^{(\rm ML)}$ is the modeling function of the ML classifier, and $\rm{predict}$ is the corresponding prediction function.

\subsection{Using Deep Neural Networks}
Long short-term memory recurrent neural network~(LSTM-RNN) and 1D-convolutional neural network~(1D-CNN) are used for predicting the lifespan of a path bin. The channel parameters of a path bin, trend changes of the channel parameters of a path bin, and the number of path bins observed during the lifespan of a given path bin are used for training. The transition probability for the death of path bins in Fig.~\ref{Fig:Markov_chain}(a) can be obtained using DNN. The channel parameters and their linear fitting of the $m^{\rm th}$ path bin used for training is given as,
\begin{equation}
    \textbf{MCP}_m = 
    \begin{bmatrix}
    {\textbf{MC}}_{m,N_m}(1) \\
    {\textbf{MC}}_{m,N_m}(2) \\
    \vdots \\
    {\textbf{MC}}_{m,N_m}(6) 
    \end{bmatrix}^{T}, 
    \qquad
    \textbf{TrLf}_m = 
    \begin{bmatrix}
    {\textbf{Tr}}_{m, N_m}(1)\\
    {\textbf{Tr}}_{m,N_m}(2) \\
    \vdots \\
    {\textbf{Tr}}_{m,N_m}(6)
    \end{bmatrix}^{T},  \label{Eq:combine}
\end{equation}
where $\textbf{MCP}_m$ represents the six dimensional channel parameter matrix of the $m^{\rm th}$ path bin over $N_m$ RX positions, the trend changes of the channel parameters of a path bin is represented by matrix $\textbf{TrLf}_m$, $\textbf{Tr}_{m,N_m}(u)$ represent the linear fitting of the $u^{\rm th}$ channel parameter of $m^{\rm th}$ path bin over $N_m$ RX positions. The number of path bins observed other than the $m^{\rm th}$ path bin during the lifespan of $m^{\rm th}$ path bin is given as $\textbf{M}_m = 
    \big[
     {M}_{m,1}^{(\rm oth)}~
     {M}_{m,2}^{(\rm oth)}~
     \cdots~
     {M}_{m,N_m}^{(\rm oth)}
    \big]^{\rm T}
    $. The overall training data matrix $\textbf{M}_m^{(\rm Tr)}$ of $m^{\rm th}$ path bin is given as $\textbf{Mat}_m^{(\rm Tr)} = [\textbf{MCP}_m~~ \textbf{TrLf}_m~~ \textbf{M}_m]$. Each data column of $\textbf{Mat}_m^{(\rm Tr)}$ is uniformly sequenced by interval $\delta n$. The data in $\textbf{Mat}_m^{(\rm Tr)}$ can be undersampled to reduce the complexity for large path bins. Similar to the ML-based prediction discussed earlier, the prediction of lifespan of a path bin is carried out using $\textbf{M}_m^{(\rm Tr)}$ and LSTM-RNN and 1D-CNN classifiers.

\subsection{Using Poisson Distribution}
If the path bins are considered independent across the RX positions, then the birth and death of path bins can be represented as state transitions, following a Markov chain. This model keeps track of the number of active path bins created by MPCs as the UAV moves, with state transitions determined by probabilities that account for MPC birth and death. These probabilities are influenced by UAV movement, scatterer distribution, and the time-varying nature of the channel. Fig.~\ref{Fig:Markov_chain}(b) shows the Markov chain model for path bin transitions, assuming small intervals where only one birth or death occurs per transition. Additionally, we assume state durations follow an exponential distribution, with transition probabilities derived from a Poisson distribution.

The birth and death of path bins can be modeled using the Poisson distribution. The birth rate and probability of birth are represented by $\nu$ and $p$, respectively, shown in Fig.~\ref{Fig:Markov_chain}(b). The probability of death of a path bin is $q$ and the death rate is given by $\mu$ shown in Fig.~\ref{Fig:Markov_chain}(b). Let $\delta n$ be the spatial step between states such that only one birth or death can occur and $\delta n = L/N$, where $L$ is the length of the UAV trajectory. 

If $y$ is the number of deaths in interval $\delta n$, we have: 
\begin{equation}
  q(y;\delta n) =
    \begin{cases}
      \mu \delta n & \text{if $y=1$}\\
      1- \mu \delta n & \text{if $y=0$}\\
      0 & \text{if $y > 1$}
    \end{cases} .      
\end{equation}
The Poisson probability mass function~(PMF) of $y$ deaths during duration $\delta n$ is given as: 
\begin{align}
    q(y;\delta n) = \frac{\big(\mu \delta n\big)^y\exp{\big(-\mu \delta n}\big)}{y!}.
\end{align}
Similarly, the PMF of the birth of path bins is obtained. The inter-arrival time of birth and death of path bins are exponentially distributed and memoryless.

\section{Simulation Results} \label{Section:simulations}
In this section, simulation setup, results, and performance of different methods for prediction of lifespan of path bins are compared. 


\begin{table}[!t]\vspace{-2mm}
	\begin{center}
		\caption{Parameters for ray tracing simulations.}\label{Table:Simulations}\vspace{-2mm}
		\resizebox{0.9\columnwidth}{!}{
        \begin{tabular}{@{} |P{5.2cm}|P{2.5cm}| @{}}
			\hline
			\textbf{Parameter}&\textbf{Parameter value}\\			
			\hline
		    Center frequency& $28$~GHz \\
            \hline
            Antenna radiation pattern & Donut \\
            \hline
            Antenna polarization& Vertical \\
            \hline
            Permittivity of ground
            &  $3.5$\\
            \hline
            Permittivity of scatterer structure
            & $5.31$ \\
            \hline
             $\gamma^{(P)}$
            & $-175.5$ \\
            \hline
            UAV velocity
            & $10$ m/s \\
            \hline
            \end{tabular}
            }
		\end{center}\vspace{-5mm}
\end{table}

 \begin{figure}[!t]
	\centering
	\includegraphics[width=\columnwidth]{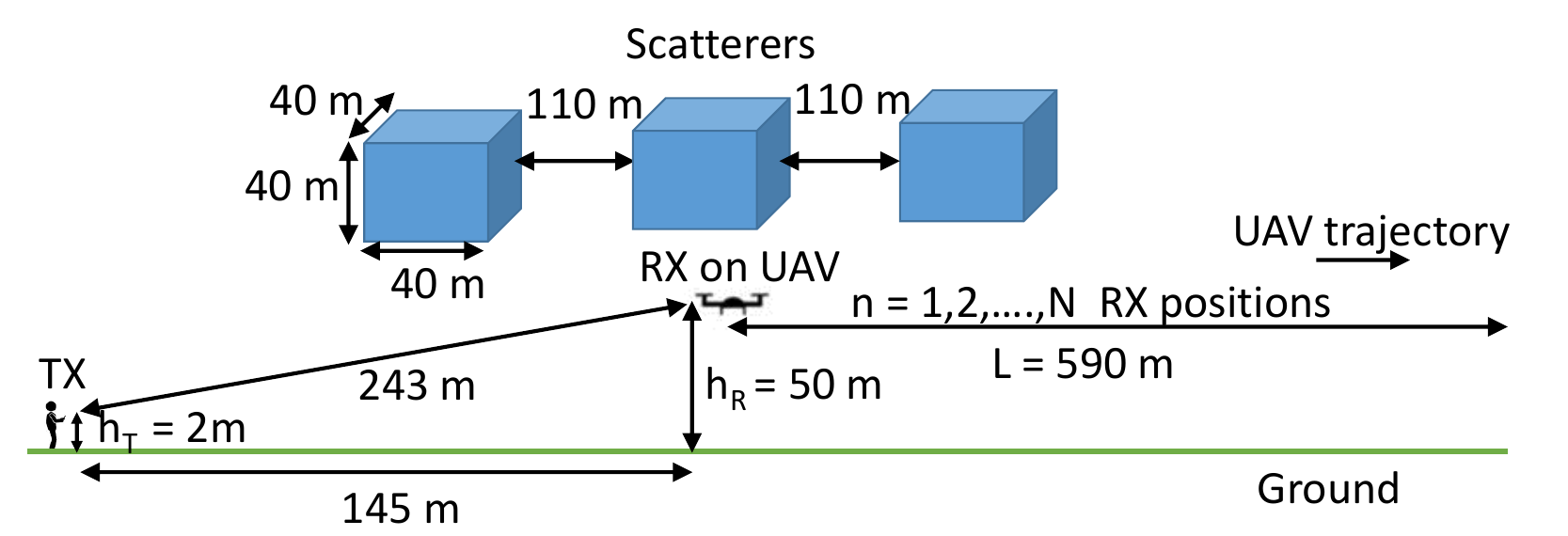}
 \captionsetup{justification=raggedright, singlelinecheck=false}
	\caption{Simulation scenario in Wireless InSite ray tracing software.    }\label{Fig:simulations_scenario}
\end{figure}

\begin{figure*}[ht!] 
  \centering
  \begin{subfigure}[b]{0.45\textwidth} 
    \centering
    \includegraphics[width=0.96\textwidth]{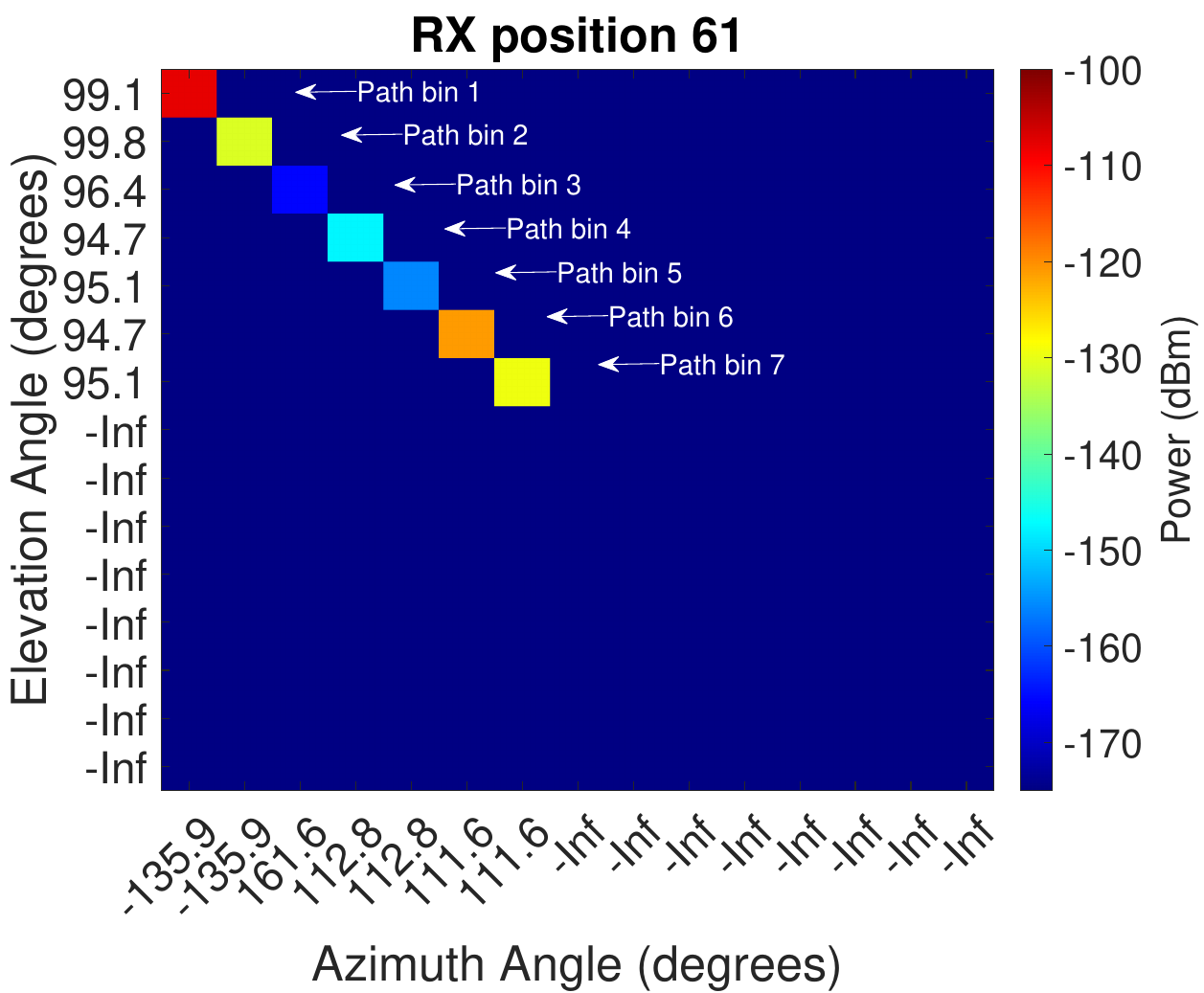} 
    \caption{ }
    \label{fig:figure1}
  \end{subfigure}
  \hfill
  \begin{subfigure}[b]{0.45\textwidth}
    \centering
    \includegraphics[width=0.96\textwidth]{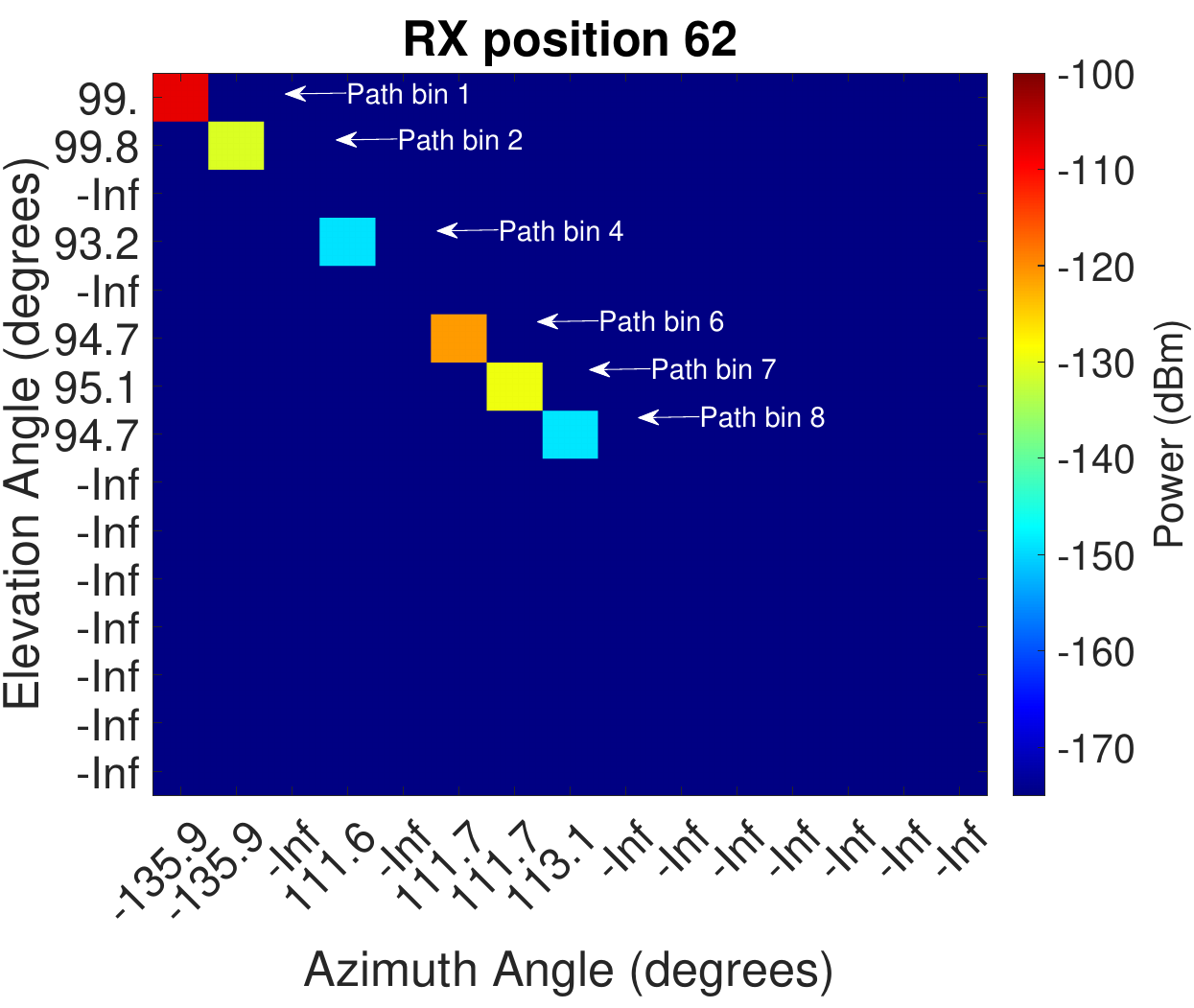} 
    \caption{}
    \label{fig:figure2}
  \end{subfigure}
  \captionsetup{justification=raggedright, singlelinecheck=false}
  \caption{Ray-based birth-death scenario at two RX locations, 61 and 62 at given elevation and azimuth angles of arrival.}
  \label{Fig:deterministic_scenario}
\end{figure*}

\subsection{Simulation Setup}
The simulations are conducted using Wireless InSite ray tracing software, which employs geometric ray-based analysis. The simulation parameters are provided in Table~\ref{Table:Simulations}. The GA propagation scenario, with scatterers positioned on one side of the TX-RX trajectory as shown in Fig.~\ref{Fig:simulations_scenario}, creates a directional propagation effect. A donut-shaped antenna pattern is applied at both the TX and RX.

We apply Monte Carlo simulations to introduce stochastic variability into ray-tracing channel data, simulating real-world randomness in the channel parameters. This approach generates diverse random samples for training and testing, preventing model overfitting and approximating real-world conditions. Testing on empirical data is proposed as future work to further validate predictive capabilities of methods. Moreover, the mean absolute error~(MAE) is used to observe the accuracy of prediction of the lifespan~(including death instance) of path bins using different methods given as $\sum_{s=1}^{S}\frac{|N_{s}^{(\rm est)} -~ N_{s}|}{S}$, where $S$ is the total number of random samples of path bins for prediction.


\begin{figure}[!t]
	\centering
	\includegraphics[width=\columnwidth]{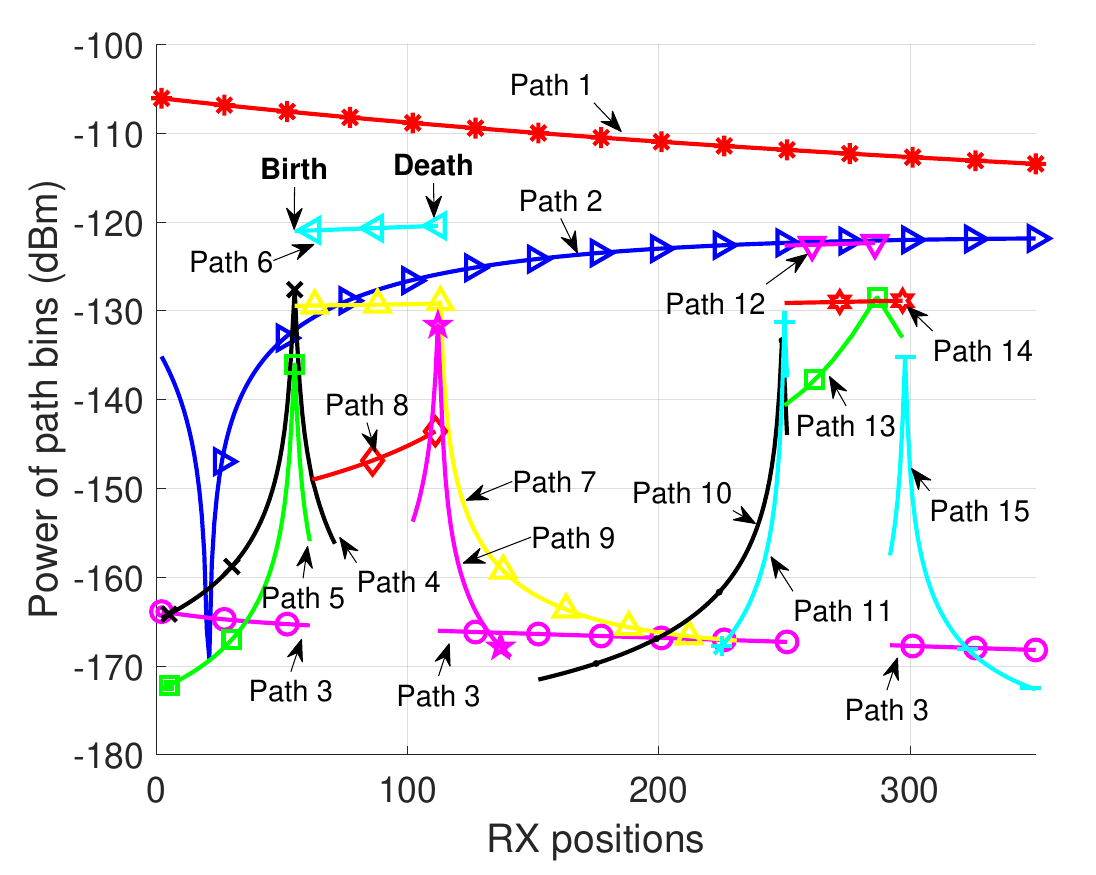}
 \captionsetup{justification=raggedright, singlelinecheck=false}
	\caption{Power of MPC path bins over the RX positions. We can observe birth and death of MPC path bins along the RX trajectory, e.g., highlighted for path bin 6.}\label{Fig:power_path_bins}
\end{figure}

\subsection{Simulation Results}
Let $\gamma^{\rm (P)}$ be the power threshold of a ray~(or MPC) required to be detected at the RX. If the received power of the ray is greater than $\gamma^{\rm (P)}$, then the birth of a path bin occurs, and if the path bin maintains that threshold over the RX positions it remains alive otherwise the death of that path bin occurs. The majority of the death of the path bins is due to a reduction in the antenna gain of the rays at TX and RX at different RX positions. The birth and death scenarios at two RX positions is shown in Fig.~\ref{Fig:deterministic_scenario}. From Fig.~\ref{Fig:deterministic_scenario}(a), for $\gamma^{\rm (P)} \geq -175.5$~dBm, we observe seven path bins at RX position 61 at respective elevation and azimuth angles. At RX position 62, Fig.~\ref{Fig:deterministic_scenario}(b) shows path bins 3 and 5 dying and path bin 8 being born.

The power of path bins as time series over RX positions obtained after applying our Euclidean distance based method is shown in Fig.~\ref{Fig:power_path_bins}. We can observe the birth and death instances, and lifespan of path bins over the RX positions. Similarly, we obtain the time series representation of other channel parameters over the RX positions.  

\begin{figure}[!t]
    \centering
    \includegraphics[width=\columnwidth]{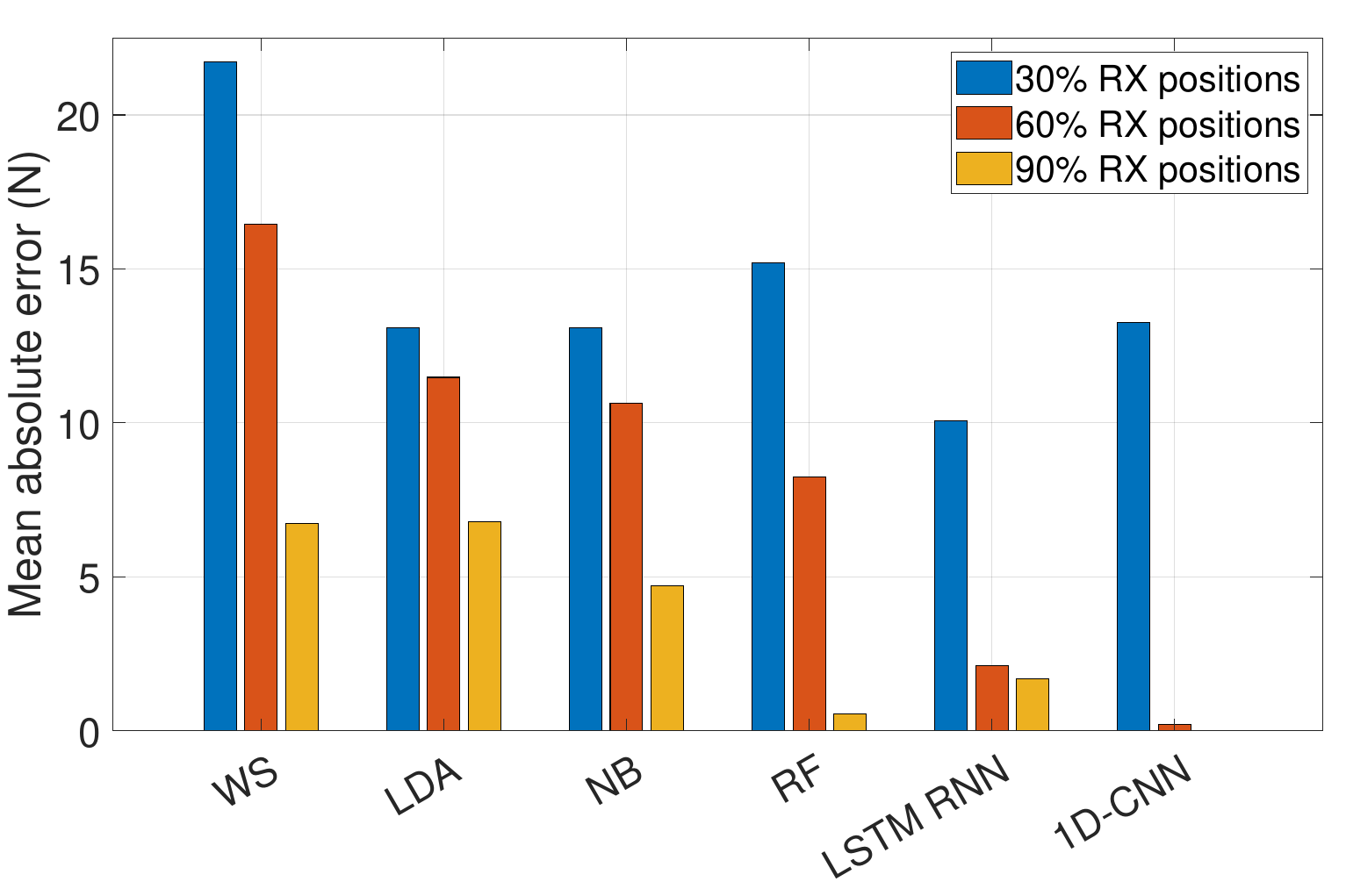}
    \caption{MAE for WS, LDA, NB, RF, LSTM-RNN, and 1D-CNN. The prediction of the death of a path bin is made at $30\%$, $60\%$, and $90\%$ of its length over RX positions.}
    \label{Fig:MAE_result}
\end{figure}

We have selected optimum tuning parameters for 1D-CNN using grid search to determine the best filter sizes, kernel sizes, and the number of filters. For RF, we used cross-validation to fine-tune the number of trees and feature set size to ensure balanced trade-offs between computational complexity and prediction accuracy. The MAE for WS, LDA, NB, RF, LSTM-RNN, and 1D-CNN are shown in Fig.~\ref{Fig:MAE_result}. The MAE in Fig.~\ref{Fig:MAE_result} is shown for path bins at $30\%$, $60\%$, and $90\%$ of their lifespan. As expected, the MAE reduces when a higher percentage of the lifespan of a path bin has already been covered. It can be observed that 1D-CNN has the best performance in terms of MAE followed by LSTM-RNN. Overall, the likelihood of death of a path bin increases when 1) the amplitude variation $\Delta \alpha$ is large, and 2) the correlation coefficient $C^{(\rm pos)}$ is very small and $C^{(\rm neg)}$ is large. 

\begin{figure}[!t]
    \centering
    \includegraphics[width=\columnwidth]{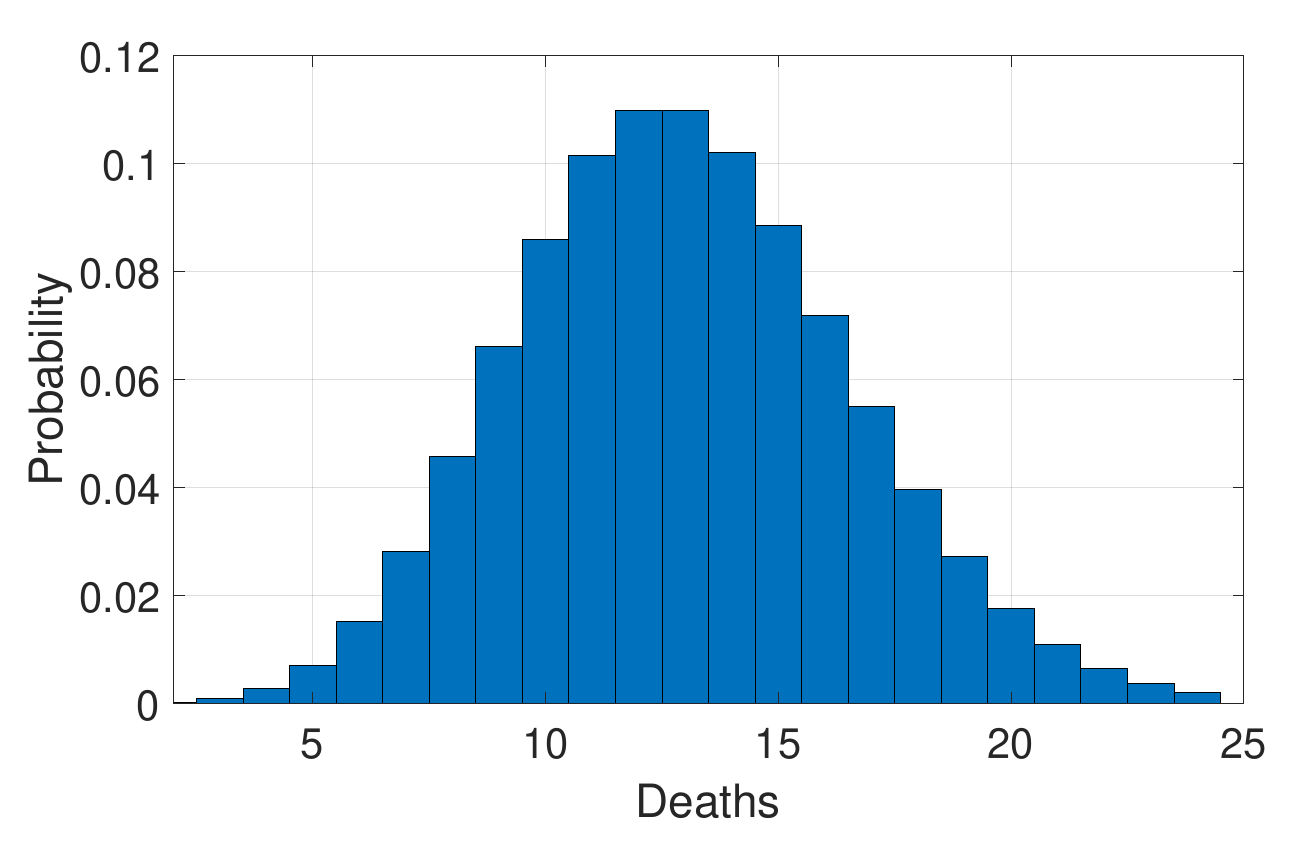}
    \caption{Poisson distribution for path bin death.}
    \label{Fig:poisson_death_dist}
\end{figure}


Poisson-based death prediction has the highest MAE of $47.2$ and is independent of the length of the path bins at the instance of prediction. Similarly, the MAE for Poisson-based birth prediction is $60.4$. The high MAE for Poission-based birth and death prediction is mainly due to the assumption of independence of path bins across RX positions and also indicating the non-WSSUS nature of the channel considered. The Poisson probability distribution of death of path bins for death rate of $\mu = 0.0371$ is shown in Fig.~\ref{Fig:poisson_death_dist}. 

The overall MAE for predicting the death of path bins at $30\%, 60\%$, and $90\%$ of their lifespans in Fig.~\ref{Fig:MAE_result} for different methods are provided in Table~\ref{Table:complexity}. Table~\ref{Table:complexity} also provides computational complexity of different methods for prediction. From Table~\ref{Table:complexity}, WS has the least computational complexity. However, its average MAE is large. LSTM-RNN has the highest computational complexity and MAE is significantly small. The average MAE and computational complexity of ML-based prediction are moderate.

\begin{table}[htbp]
	\begin{center}
     \footnotesize
		\caption{Computational complexity and average MAE for prediction of the death of path bins using different methods. The unit of MAE is represented by $N$ (number of RX positions). Here $e$ is the number of events, $n_1^{(\rm f)}$, $n_2^{(\rm f)}$  is the number of features, and $n_1^{(\rm f)}=9$ and $n_2^{(\rm f)}=13$, $n^{(\rm s)}=100$, are the number of times the path bins are randomly sampled, $n^{(\rm tr)}=100$, is the number of trees, $n^{(\rm h)}=120$, is the number of hidden units, $n^{\rm (fil)}=32$, is the number of filters, and $n^{(\rm ker)}=5$ represents the kernel size.  } \label{Table:complexity}
\begin{tabular}{@{}|P{1.05cm}|P{5.3cm}|P{1.25cm}|@{}}
 \hline
\textbf{Predictor}&\textbf{Computational complexity}&\textbf{
Overall MAE ($N$)}\\
\hline
Poisson&$O\big(e\big)$&$47.2$\\
\hline
WS&$O\big(M(n^{\rm (f)}_{1})^2\big)$&$14.97$\\
\hline
LDA & $O\Big(M n^{(\rm s)}(n_1^{(\rm f)})^3 + (n_1^{(\rm f)})^3 \Big)$ &$10.45$ \\
\hline
NB& $O\big(M n^{(\rm s)} (n^{(\rm f)}_{1})^2\big)$ & $9.48$ \\
\hline
RF& $O\big(M n^{(\rm s)} n^{(\rm f)}_{1}\log\big(M n^{(\rm s)} n^{(\rm f)}_{1}\big) n^{(\rm f)}_{1} n^{(\rm tr)}\big)$& $8$\\
\hline
LSTM-RNN & $O\Big(n^{(\rm h)}\big(4\big(n^{(\rm f)}_{2}\big)^2 N + 4n^{(\rm h)} n^{(\rm f)}_{2} N + 4n^{(\rm f)}_{2} N\big)\Big)$ 
&$4.62$\\
\hline 1D-CNN & $O\Big(n^{(\rm f)}_{2}n^{(\rm fil)}n^{(\rm ker)}\big(n^{(\rm f)}_{2} N - n^{(\rm ker)} + 1 \big)\Big)$ & $4.5$\\
\hline
\end{tabular}
		\end{center}
			\end{table}

The prediction of the birth and death of MPCs using Poisson distribution has the major advantages of very low computational complexity, simplicity, and scalability. However, the prediction results suffer in highly variable, dependent, and clustered event scenarios. The WS is also simple and scalable, however, the prediction accuracy is dependent on the optimal selection of the weights. Furthermore, the assumption of linearity and not considering the interactions among the variables can lead to inaccuracies in the prediction process. Both ML and DNN are computationally extensive and require training, however, they are adaptive to diverse situations leading to high prediction accuracy. DNN has higher accuracy compared to ML methods due to a larger number of hidden layers and the ability to manage, and extract features from unstructured data.

\section{Conclusions and Future Work}  \label{Section:conclusions}
In this work, we have considered the UAV GA propagation channel as non-WSSUS and used the correlation among the MPCs to create path bins and predict their lifespan. WS, ML, and DNN methods are used to predict the lifespan of the path bins by using the correlation among the path bins across the RX positions and channel parameter variations. Poisson distribution and Markov chain is also used to model and predict the birth and death of MPC path bins over RX positions. The prediction results indicate that the Poisson-based method has the least computational complexity, however, MAE is the highest. The WS method has lower computational complexity compared to ML and DNN, however, the MAE is large. The DNN has higher complexity compared to other methods but the lowest MAE indicating an accurate prediction of the death of path bins. The scalability of deep learning methods to large datasets, real-time prediction scenarios, and their applicability in resource allocation for UAV communications will be explored in the future work.

\balance 

\bibliographystyle{IEEEtran}
\bibliography{References}

\begin{thebibliography}{10}
\providecommand{\url}[1]{#1}
\csname url@samestyle\endcsname
\providecommand{\newblock}{\relax}
\providecommand{\bibinfo}[2]{#2}
\providecommand{\BIBentrySTDinterwordspacing}{\spaceskip=0pt\relax}
\providecommand{\BIBentryALTinterwordstretchfactor}{4}
\providecommand{\BIBentryALTinterwordspacing}{\spaceskip=\fontdimen2\font plus
\BIBentryALTinterwordstretchfactor\fontdimen3\font minus
  \fontdimen4\font\relax}
\providecommand{\BIBforeignlanguage}[2]{{%
\expandafter\ifx\csname l@#1\endcsname\relax
\typeout{** WARNING: IEEEtran.bst: No hyphenation pattern has been}%
\typeout{** loaded for the language `#1'. Using the pattern for}%
\typeout{** the default language instead.}%
\else
\language=\csname l@#1\endcsname
\fi
#2}}
\providecommand{\BIBdecl}{\relax}
\BIBdecl

\bibitem{uav_AG_ref1}
X.~Lin, V.~Yajnanarayana, S.~D. Muruganathan, S.~Gao, H.~Asplund, H.-L.
  Maattanen, M.~Bergstrom, S.~Euler, and Y.-P.~E. Wang, ``The sky is not the
  limit: {LTE} for unmanned aerial vehicles,'' \emph{IEEE Commun. Mag.},
  vol.~56, no.~4, pp. 204--210, 2018.

\bibitem{uav_AG_ref2}
W.~Khawaja, I.~Guvenc, D.~W. Matolak, U.-C. Fiebig, and N.~Schneckenburger, ``A
  survey of air-to-ground propagation channel modeling for unmanned aerial
  vehicles,'' \emph{IEEE Commun. Surv. \& Tut.}, vol.~21, no.~3, pp.
  2361--2391, 2019.

\bibitem{uav_5G_beyond1}
B.~Li, Z.~Fei, and Y.~Zhang, ``{UAV} communications for {5G} and beyond: Recent
  advances and future trends,'' \emph{IEEE Internet of Things J.}, vol.~6,
  no.~2, pp. 2241--2263, 2018.

\bibitem{uav_5G_beyond2}
Y.~Zeng, Q.~Wu, and R.~Zhang, ``Accessing from the sky: A tutorial on {UAV}
  communications for {5G} and beyond,'' \emph{Proc. of the IEEE}, vol. 107,
  no.~12, pp. 2327--2375, 2019.

\bibitem{3GPP_ntn}
{Third Generation Partnership Project (3GPP)}, Technical Report, TR 21.918,
  Release 18, version 0.2.0, Feb., 2024.

\bibitem{birth_death1}
Z.~Zhang, Y.~Liu, J.~Huang, J.~Zhang, J.~Li, and R.~He, ``Channel
  characterization and modeling for {6G UAV}-assisted emergency communications
  in complicated mountainous scenarios,'' \emph{Sensors}, vol.~23, no.~11, p.
  4998, 2023.

\bibitem{birth_death2}
Z.~Lian, L.~Jiang, C.~He, and Q.~Xi, ``A novel multiuser {HAP-MIMO} channel
  model based on birth-death process,'' in \emph{Proc. IEEE Int. Conf. Commun.
  (ICC)}, 2016, pp. 1--5.

\bibitem{wahab_AG_previous}
W.~Khawaja, O.~Ozdemir, and I.~Guvenc, ``Channel prediction for {mmWave}
  ground-to-air propagation under blockage,'' \emph{IEEE Ant. and Wireless
  Propag. Lett.}, vol.~20, no.~8, pp. 1364--1368, 2021.

\bibitem{birth_death_tracking1}
D.~Méndez-Romero and M.~J. F.-G. García, ``Simpler multipath detection for
  vehicular {OFDM} channel tracking,'' \emph{IEEE Trans. Vehic. Technol.},
  vol.~67, no.~11, pp. 10\,752--10\,759, 2018.

\bibitem{birth_death_tracking2}
M.~L. Jakobsen, T.~Pedersen, and B.~H. Fleury, ``Analysis of stochastic radio
  channels with temporal birth-death dynamics: A marked spatial point process
  perspective,'' \emph{IEEE Trans. Ant. and Propag.}, vol.~62, no.~7, pp.
  3761--3775, 2014.

\bibitem{birth_death_tracking4}
Z.~Huang, J.~Rodríguez-Piñeiro, T.~Domínguez-Bolaño, X.~Cai, and X.~Yin,
  ``Empirical dynamic modeling for low-altitude {UAV} propagation channels,''
  \emph{IEEE Trans. Wireless Commun.}, vol.~20, no.~8, pp. 5171--5185, 2021.

\bibitem{BD_Matolak}
D.~W. Matolak, ``Air-ground channels \& models: Comprehensive review and
  considerations for unmanned aircraft systems,'' in \emph{Proc. IEEE Aerospace
  Conf.}, 2012, pp. 1--17.

\bibitem{non-wssus}
L.~Bernadó, T.~Zemen, F.~Tufvesson, A.~F. Molisch, and C.~F. Mecklenbräuker,
  ``The (in-) validity of the {WSSUS} assumption in vehicular radio channels,''
  in \emph{Proc. IEEE Int. Symp. on Pers., Indoor and Mobile Radio Commun. -
  (PIMRC)}, 2012, pp. 1757--1762.

\bibitem{overdet_norm}
L.~Richard~Jr \emph{et~al.}, \emph{Scientific Data Analysis: an introduction to
  overdetermined systems}.\hskip 1em plus 0.5em minus 0.4em\relax Springer
  Science \& Business Media, 2012.

\end{thebibliography}

\end{document}